\documentclass[preprint,floats,aps,nofootinbib,amssymb]{revtex4}
\usepackage{mathrsfs}
\usepackage{slashed}
\usepackage{graphicx}
\usepackage{color}
\usepackage{dcolumn}
\usepackage{bm}
\usepackage{graphicx}
\usepackage{amssymb}
\usepackage{stackrel}

\begin{document}

\title{ The electron-flavored $ \mathbf{Z}^\prime$ portal dark matter and the  DAMPE  cosmic ray excess  }

\author{Wei Chao$^1$}
\email{chaowei@bnu.edu.cn}
\author{Qiang Yuan$^{2,3}$}
\email{yuanq@pmo.ac.cn}
\affiliation{$^1$Center for Advanced Quantum Studies, Department of Physics, Beijing Normal University, Beijing, 100875, China\\
$^2$Key Laboratory of Dark Matter and Space Astronomy, Purple Mountain Observatory, Chinese Academy of Sciences, Nanjing 210008, China\\
$^3$ School of Astronomy and Space Science, University of Science and Technology of China,  Heifei, Anhui, 230026, China}

\vspace{3cm}

\begin{abstract}
The DAMPE experiment has recently reported strong indications for the existence of an excess of high-energy electrons and positrons.
If  interpreted in terms of the annihilation of dark matter, the DAMPE result restricts the dark matter mass and possible annihilation channels to a few case.
In this paper we explain the DAMPE result with the electron-flavored $Z^\prime$-portal  fermionic dark matter. 
We show that the Dirac dark matter scenario is promising to explain the excess via the process $\bar  \chi \chi \to\mathbf{Z}'\to \bar e e$.
The reduced annihilation cross section is limited in a range of $10^{-26}\sim 10^{-24}~{\rm cm^3 s^{-1}}$ to interpret the excess.

\end{abstract}

\maketitle

\section{Introduction}
Astrophysical evidence shows that about 80\% of the mass in the Universe is made of non-luminous and non-baryonic matter~\cite{Zwicky:1933gu,Ade:2015xua,Aubourg:2014yra}, but the nature of the particle that constitutes this dark matter and the way it interacts with the Standard Model (SM) particles remain mysteries, which catalyzes various portals of the dark matter and extensions to the minimal SM.
Of various dark matter models the Weakly Interacting Massive Particle (WIMP) is promising, since the WIMP relic abundance can be naturally around the experimental value, $\Omega h^2 =0.1189\pm0.0023$~\cite{Patrignani:2016xqp} for a WIMP mass around the electroweak scale. 
There are a variety of on-going and future experiments searching for WIMP, including direct detection experiments, which aim to observe the elastic WIMP-nuclei scattering cross section, and  indirect detection experiments, which aim to observe cosmic rays from the annihilation or decay of WIMP in the galaxy.

Although no signal in direct detection experiments has been observed up to now, which puts upper-bound  on the WIMP-nucleus scattering cross section,  an indirect detection experiment, Dark Matter Particle Explorer (DAMPE; \cite{ChangJin:550,TheDAMPE:2017dtc}), brings dawn to the identification of dark matter, which observed  an excess in the electron/positron cosmic-ray  spectrum at around $1.5$ TeV~\cite{dampe-data}. 
There are already some interpretation to the excess~\cite{Yuan:2017ysv,Fan:2017sor,Fang:2017tvj,Duan:2017pkq,Gu:2017gle,Zu:2017dzm,Tang:2017lfb,Huang:2017egk,Athron:2017drj,Cao:2017ydw,Liu:2017rgs,Chao:2017emq,Duan:2017qwj,Gu:2017bdw}. 
In this paper we propose a dark matter explanation to this cosmic ray excess  from the bottom-up approach: the electron-flavored $Z^\prime$-portal dark matter model.
The model extends the SM with a $U(1)_\mathbf{X}^{}$ gauge symmetry, where only the right-handed electron  of the SM particles carries non-zero $U(1)_\mathbf{X}^{}$ charge.
Anomalies can be eliminated by introducing a vector-like electroweak singlet. 
We investigate constraint on the model from the LEP experiment, which has $M_{\mathbf{Z}'}/g_\mathbf{X} >2.42~{\rm TeV}$ with $g_\mathbf{X}$ the new gauge coupling and $M_\mathbf{Z'}$ the mass of the new gauge boson.
We further show that the electron mass can be generated from a seesaw mechanism.  

The dark matter $\chi$, which carries unit $U(1)_\mathbf{X}^{}$ charge, can be Dirac or Majorana fermion stabilized by the $U(1)_\mathbf{X}^{}$.
If $\chi$ is a Dirac fermion, there is no direct detection constraint on the model since $\mathbf{Z'}$ is leptonphilic  and the DAMPE cosmic ray excess can be explained from the annihilation of the dark matter into electron pairs directly in a sub-structure of the dark halo. 
It shows that the reduced annihilation cross section is limited in a range of $10^{-26}\sim 10^{-24}~{\rm cm^3 s^{-1}}$ to interpret the excess.
If $\chi$ is a Majorana fermion, the $\chi-nucleus$ cross section is proportional to  $\sin^2 2 \alpha$, where  $\alpha$ is the mixing angle between the SM Higgs and the extra scalar singlet,  and direct detection constraints can be evaded by a small mixing angle, which is favored by the electroweak precision measurements. 
Electrons and positrons may come from the process $\chi \bar \chi \to \mathbf{Z}'\mathbf{Z}'$, which subsequently decay into $e\bar e e \bar e$ with fraction almost $100\%$, but it can hardly fit the observed cosmic ray excess considering the constraint of the LEP.
%

The resulting of the paper is organized as follows: in section II we introduce the model in detail.
In section III we study the dark matter relic abundance and discuss possible constraint from direct detection. 
Section IV is focused on the DAMPE cosmic ray excess.
The last part is concluding remarks.

\section{The Model }

\begin{table}[t]
\begin{tabular}{c|c|c||c|c|r}
\hline
\hline SM particles &$G_{\rm GM}$ & $U(1)_\mathbf{X}^{} $  & Beyond SM particles & $G_{\rm SM}$ & $U(1)_\mathbf{X}^{}$ \\
\hline
$ \ell_L^i$ &($1$, $2$, $-{1\over 2}$)  & 0 & $ \psi_L^{} $ & $(1, ~1, -1)$ & $2$\\
$e_R^{}$ & $(1,~1, -1)$ & 2 &  $ \psi_R^{} $ & $(1,~1, -1)$ & $0$  \\
$Q_L^i$& $(3,~2,~{1\over 6})$ & 0 & $\chi_R$&(1,~1,~0)  &$1$ \\
$U_R^i$& $(3,~1,~{2\over 3})$&0&$\chi_L $ & (1,~1,~0) & $1$ \\
$D_R^i$& $(3,~1,~{1\over 3})$&0&$S$ & (1,~1,~0) & $2$\\
\hline
$H$ &$(1,~2,~{1\over 2})$& 0 \\
$\mu_R^{}$ & $(1,~1, -1)$ & 0 \\
$\tau_R^{}$ & $(1,~1, -1)$ & 0 \\
\hline 
\hline
\end{tabular}
\caption{ Quantum numbers of various fields under the local $U(1)_{\mathbf{X}}$, $\ell_L^i$ and $Q_L^i$ $(i=1,2,3)$ are left-handed lepton doublet and quark doublet, $U_R^i$ and $D_R^i$ are right-handed up-type quark and down-type quark respectively.  }\label{aaa}
\end{table}

Many models with new U(1) gauge symmetry have been proposed addressing various problems, such as $U(1)_{\mathbf{L}_i-\mathbf{L}_j}$~\cite{He:1991qd}, $U(1)_{\mathbf{N}}$~\cite{King:2005jy},  $U(1)_{\mathbf{ B-L}}$~\cite{Mohapatra:1980qe,Marshak:1979fm,Wetterich:1981bx}, $U(1)_\mathbf{B}$~\cite{FileviezPerez:2010gw,Dulaney:2010dj}, $U(1)_{\mathbf{L}}$~\cite{FileviezPerez:2010gw,Dulaney:2010dj,Chao:2010mp}, $U(1)_{\mathbf{B+L}}$~\cite{Chao:2015nsm,Chao:2016avy}, generic U(1)~\cite{Appelquist:2002mw,Ekstedt:2016wyi}, $U(1)_\mathbf{R}$~\cite{Chao:2017rwv} etc.
For a review of various U(1) models and collider signatures of the U(1)-related gauge boson, we refer the reader to Ref.~\cite{Langacker:2008yv} for detail. 
The $Z'$ gauge boson in these models may arise as the mediator between dark and visible sectors.
Robust constraints may arise form collider, direct and indirect searches~\cite{Alves:2013tqa,Alves:2016cqf,Arcadi:2017hfi}, as well as the muon magnetic moment.
Of various $Z^\prime$ models the leptonphelic U(1) model is popular since it may avoid constraints of direct detection. 
In this section we introduce an electron-flavored U(1) model where only right-handed electron of the SM particles carries non-zero charge.
One vector like fermion $\psi_{L,R}^{}$ is introduced to cancel anomalies of the model.
Particle contents and their corresponding charges are listed in the table.~\ref{aaa}, where $\chi$ is the dark matter candidate stabilized by the $U(1)_\mathbf{X}$.
It is easy to check that all potential anomalies, i.e., the axial-vector anomalies~\cite{avector1,avector2,avector3}  ${\cal A}_1 (SU(2)_L^2\otimes U(1)_Y^{}  )$, ${\cal A}_2 (U(1)_Y^2 \otimes U(1)_L^{}  )$, ${\cal A}_3 (U(1)_L^2 \otimes U(1)_Y^{}  )$,~${\cal A}_4 (U(1)_L^{3}  )$ and the gravitational-gauge anomaly~\cite{anog1,anog2,anog3} ${\cal A}_5 ( U(1)_L^{}  )$, are automatically canceled in this simple framework, which are listed in the table.~\ref{anomaly} in detail.

\begin{table}[t]
\centering
\begin{tabular}{l|l}
\hline anomalies~~~~~~~~~~~~~~ & anomaly free conditions~~~~~~~~~~~~~~ \\
\hline
$U(1)_Y^2 U(1)_{\mathbf{X}}^{} $: & $ (-1)^2\times (2-2)=0$ \\
$U(1)_{\mathbf{X}}^2 U(1)_Y$:&  $2^2\times [(-1)-(-1)]=0 $ \\
$U(1)_{\mathbf{X}}$: & $ 1-1+(2-2)=0 $ \\
$U(1)_{\mathbf{X}}^3$: &  $1^3-1^3+2^3-2^3=0$ \\
\hline
\end{tabular}
\caption{ The anomaly cancellation conditions of the $U(1)_{\mathbf{X}}$.  }\label{anomaly}
\end{table}

\subsection{The scalar sector}
The general Higgs potential takes the form
\begin{eqnarray}
V=-\mu_h^2 H^\dagger H +\lambda(H^\dagger H)^2 +\lambda (H^\dagger H)^2 - \mu_s^2 S^\dagger S -\lambda_1^{} (S^\dagger S) + \lambda_2^{} (H^\dagger H) (S^\dagger S)
\end{eqnarray}
where $H=(H^+ ,(v_h+h+iG)/\sqrt{2})^T$ and $S=( s+iG'+v_s)/\sqrt{2}$ with $v_h$, $v_s$ the vacuum expectation values of $H$ and $S$ respectively.
There are two CP-even scalars after spontaneous symmetry breaking with mass eigenvalues 
\begin{eqnarray}
M_{1,2}^2 = (\lambda v_h^2 + \lambda_1 v_s^2 ) \pm \sqrt{ (\lambda v_h^2 - \lambda_1 v_s^2 )^2 + \lambda_2^2 v_h^2 v_s^2  },
\end{eqnarray}
where the smaller one is taken as the mass square of the SM Higgs.  
The mixing angle ($\alpha$) between $h$ and $s$ can be calculated analytically with $\tan^{-1} 2\alpha =-(\lambda_1^{} v_s^2 -\lambda v_h^2 )/(\lambda_2^{} v_s^{} v_h^{})$.
The size of $\alpha$ is constrained by the oblique observables as well as the Higgs measurement at the LHC. 
We refer the reader to Refs.~\cite{Profumo:2014opa,Chao:2016vfq,Chao:2016cea} for the study of these constraints in detail.  

\subsection{The $Z^\prime$ mass}
The mass of $Z^\prime$ can be written as 
\begin{eqnarray}
M_{\mathbf{Z}'}^{} = 2 g_\mathbf{X}^{} v_s^{} 
\end{eqnarray}
where $g_\mathbf{X}$ is the gauge  coupling of $U(1)_{\mathbf{X}}$. 
The $\mathbf{Z'}^{}$ mass may be constrained by colliders. 
Since $\mathbf{Z'}^{}$ does not couple to quarks, there is no constraint from the LHC and we only consider constraint from the LEP, where the combined leptonic $e^+e^-\to \ell \bar \ell$ cross sections as well as the leptonic forward-background asymmetries are used to fit the data to models including neutral bosons.
Following the Ref. \cite{Eichten:1983hw}, the $\mathbf{Z'}^{}$ mediated interactions are parametrized by an effective Lagrangian ${\cal L}_{\text{eff}}$, which takes the form: 
\begin{eqnarray}
{\cal L}_{\text{eff} }^{}= {4\pi \over (1+\delta ) \Lambda^2   } \bar e_R^{} \gamma_\mu e_R^{}  \bar e_R^{}  \gamma^\mu e_R^{} 
\end{eqnarray}
where $\delta =0(1)$ for $\ell\neq e(\ell=e)$ and $\Lambda = \sqrt{4\pi} M_{\mathbf{Z'}}/g_\mathbf{L}$.
Fitting with the 95\% confidence limits on the scale $\Lambda $ measured at the LEP, as shown in the table 3.15 of the Ref.~\cite{Schael:2013ita}, one has $M_{\mathbf{Z'}}/g_\mathbf{L}\geq 2.42~\text{TeV}$, which results in $v_s \geq 1.21~\text{TeV}$.

\subsection{The electron mass}
Due to the $U(1)_\mathbf{X}^{}$ symmetry, right-handed electron does not couple to the lepton doublet and the SM Higgs. 
Interactions relevant to the electron and new charged fermion masses can be written as
\begin{eqnarray}
-{\cal L}_{\text{mass}} =y_E^{} \overline{\ell_E} H \psi_R^{} +y_\psi^{}  \overline{\psi_R^{} } S \psi_L^{} + m' \overline{\psi_L^{} } E_R^{}  + {\rm h.c.}
\end{eqnarray}
The effective operator for electron mass may arise from integrating out $\psi_{L,R}^{}$. 
One can also get electron mass from the diagonazition of following $ 2 \times 2 $ mass matrix  ${\cal M}$:
\begin{eqnarray}
\overline{\left( \matrix{e_L & \psi_L}\right ) } \left( \matrix{ 0& y_E^{} v/\sqrt{2} \cr m'&  y_\psi v_s/\sqrt{2} } \right)  \left( \matrix{ E_R \cr \psi_R^{} } \right) + {\rm h.c.} 
\end{eqnarray}
${\cal M}$ can be diagonalized as ${\cal U}_L^\dagger {\cal M} {\cal U}_R^{} ={\rm diag} (m_e, ~m_\psi)$, where ${\cal U}_L^{}$ and ${\cal U}_R^{} $ are $2\times 2$ unitary transformation. 
The electron and the new charged fermion masses will be $m_e \approx m^{'}y_E v_h^{} / (y_\psi v_s )$, $m_\psi \approx \sqrt{y_\psi^2  v_s^2/2 + y_E^2 v_h^2 /2 + m^{'2} }$.
The unitarity constraint~\cite{Antusch:2006vwa} on the model can be easily satisfied, which is similar to the case of the type-III seesaw mechanism~\cite{Abada:2007ux}.

\section{Dark Matter}

We will focus on the dark matter phenomenology in this section.  Interactions relating to  new neutral  fermions can be written as
\begin{eqnarray}
-{\cal L}_{\rm Y} =  {1\over 2 } \kappa_1^{} \overline{\chi_L^{} } S \chi_L^C + {1\over 2 } \kappa_2^{} \overline{\chi_R^C } S \chi_R^{} + m \overline{\chi_L^{} } \chi_R^{} + {\rm h.c.} \; ,
\end{eqnarray}
There are two scenarios corresponding to the Dirac or Majorana nature of the dark matter:  scenario I: $\kappa_1=\kappa_2 =0$, the dark matter is a Dirac fermion;  scenario II: $\kappa_1$ and(or) $\kappa_2\neq0$, the dark matter is Majorana particle.  
we will discuss these two scenarios separately in the following.
\subsection{Scenario I}

For this case, the dark matter is a Dirac fermion: $\chi=\chi_L + \chi_R$ and only interact with the $\mathbf{Z}'$: $g_\mathbf{X} \bar \chi \gamma_\mu^{} \chi \mathbf{Z}^{\prime}_\mu$.  
To estimate the relic density of the dark matter, one needs to evaluate their annihilation cross sections. 
The thermal averaged annihilation cross section for the $\bar \chi \chi \to e^+ e^-$ process can be written as 
\begin{eqnarray}
&&\langle \sigma_{\bar \chi \chi \to e \bar e } v\rangle =  {2 g_\mathbf{X}^4 m^2 \over \pi (M_{\mathbf{Z}'}^2 -4 m^2 )^2 + M_{\mathbf{Z'}}^2 \Gamma_{\mathbf{Z}'}^2}  + { g^4_\mathbf{X} m^2 (M_{\mathbf{Z}'}^2 -28m^2 ) \over 6\pi [(M_{\mathbf{Z}'}^2 -4 m^2 )^2 +M_{\mathbf{Z'}}^2 \Gamma_{\mathbf{Z}'}^2 ]^{3/2}} \langle v^2 \rangle  \label{anniA}
\end{eqnarray}
where $v$ is the relative velocity of annihilating dark matter and we have neglected the electron mass. 
The $\Gamma_{\mathbf{Z}'}$ in the denominator of Eq. (\ref{anniA}) is the decay width $\mathbf{Z}'$, which can be written as 
\begin{eqnarray}
\Gamma_{\mathbf{Z}'} =\sum_f {g_\mathbf{X}^2  M_{\mathbf{Z}'}\over 12 \pi } \sqrt{ 1 - {4 m_f^2 \over M_{Z'}^2}} \left( 1- {m_f^2 \over M_{Z'}^2 }\right) \; .
\end{eqnarray}
where $f=e$ (or $\psi$).

For the case $m>M_{\mathbf{Z}{'}} $,  dark matter will annihilate directly into pairs of on-shell  $\mathbf{Z}'$.
The non-relativistic form of  the thermal averaged  reduced annihilation cross section can be written as
\begin{eqnarray}
&&\langle \sigma_{\chi \bar \chi \to \mathbf{Z'Z'} } v \rangle  =  {g_\mathbf{X}^4 \over 4 \pi m^2 } \left( 1- {M_{\mathbf{Z'}}^2 \over m^2 } \right)^{3/2}  \left( 1- {M_{\mathbf{Z}'}^2 \over 2m^2 } \right)^{-2} + \nonumber \\
&& { \langle v^2\rangle  g_\mathbf{X}^4 M_{\mathbf{Z}'}^2 \over 64 \pi m^4} \left( 1- {M_{\mathbf{Z}'}^2 \over m^2 } \right)^{1/2}  \left( 1- {M_{\mathbf{Z}'}^2 \over 2m^2 } \right)^{-4} \left( 76 -66 {m_{\mathbf{Z}'}^2 \over m^2 } + 23 { M_{\mathbf{Z}'}^4 \over m^4 }  \right) \; .
\end{eqnarray}
%

\begin{figure*}[t]
\includegraphics[width=0.49\textwidth]{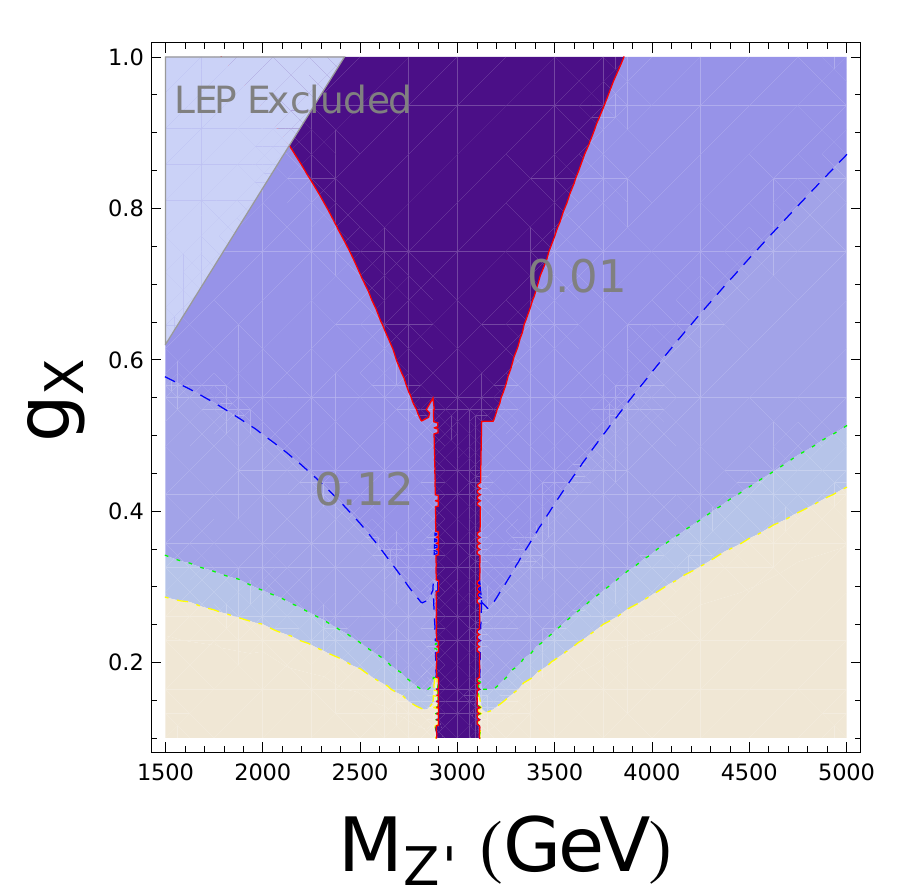}
\includegraphics[width=0.49\textwidth]{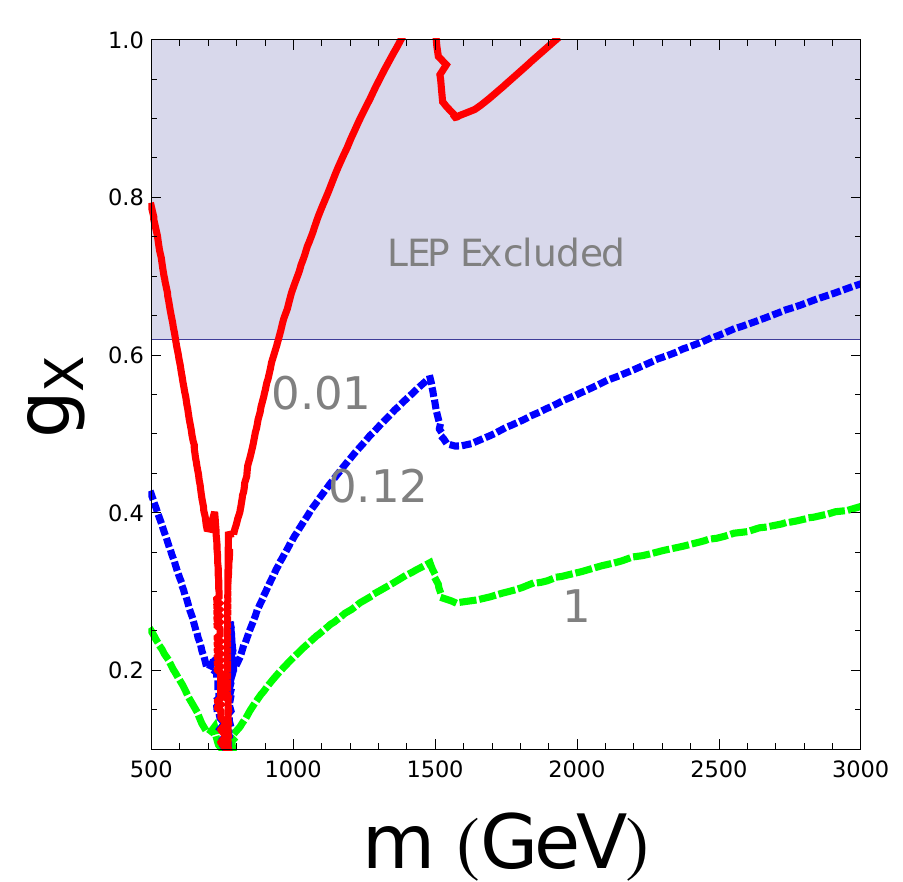}
\caption{Left-panel: contours of of the dark matter relic density in the $M_{\mathbf{Z}'}-g_\mathbf{X}$ plane by setting $m=1.5~\text{TeV}$, the region in the top-left is excluded by the LEP; Right-panel: contours of the dark matter relic density in the $m-g_\mathbf{X}$ plane by setting $M_{\mathbf{Z}'}=1.5~\text{TeV}$, the region  above the horizontal line  is excluded by the LEP.
\label{fig:relicA}}
\end{figure*}

Given these expression, one can evaluate the relic density of the dark matter, which can be written as 
\begin{eqnarray}
\Omega h^2 \approx 2 \times  {1.07 \times 10^9 \text{GeV}^{-1} \over M_{pl}^{} } {x_F \over \sqrt{g_*} } {1 \over a + 3 b/x_F} 
\end{eqnarray}
where the factor $2$ on the right side comes from the fact that the dark matter is Dirac fermion, $M_{pl} =1.22\times 10^{19}~\text{GeV}$ being the plank mass, $g_*$ is the effective degrees of freedom at the freeze-out temperature, $x_F =m/T_F$ with $T_F$ the freeze-out temperature, $a$ and $b$ are the s-wave and p-wave parts of the reduced annihilation cross section respectively.  
There are only three free parameters in the expression of the relic density: $g_\mathbf{X}$, $M_{\mathbf{Z}'}$ and $m$.

For illustration, we show in the left-panel of the Fig.~\ref{fig:relicA} contours of the dark matter relic density in the $M_{\mathbf{Z}'}-g_\mathbf{X}$ plane by setting $m=1.5~\text{TeV}$. 
The solid, dashed and dotted contours correspond to $\Omega h^2 =0.01, 0.12$ and $1$ respectively. 
The region on the top-left is excluded by the LEP. 
$M_{\mathbf{Z}'}=3~\text{TeV}$ corresponds to the resonance regime of the dark matter annihilation. 
We show in the right-panel of the Fig.~\ref{fig:relicA} contours of the dark matter relic density in the $m-g_\mathbf{X}$ plane by setting $M_\mathbf{Z'}=1.5~\text{TeV}$. 
The solid, dotted and dashed lines correspond to $\Omega h^2 =0.01,~0.12$ and $1$ respectively.  
Apparently the observed dark matter relic abundance can be explained in this model. 

For  the direct detection, considering that $\chi$ does not couple to the scalar sector and $\mathbf{Z}'$ only interact with leptons, there is no spin-independent and (or) spin-dependent direct detection signals in the tree level. 
Possible signals may come from indirect detections, which will be studied in the next section.

\subsection{Scenario II }

In this scenario the dark matter is a Majorana particle.  
After the spontaneous breaking of the electroweak and the $U(1)_\mathbf{X}$ symmetries, new neutral fermions get non-zero masses and the lightest one will be stable dark matter candidate, which is the mixture of  $\chi_{L,R}$  and $\chi_{L,R}^C$.
The new fermion mass matrix can be written as
\begin{eqnarray}
{\cal L}_{\rm mass} ={1\over 2 } \overline{(\chi_L, ~\chi_R^C)} \left( \matrix{ \kappa_1 v_s /\sqrt{2}  &  m   \cr  m  & \kappa_2  v_s /\sqrt{2}}\right) \left(\matrix{ \chi_L^C \cr \chi_R}\right)+ {\rm h.c.}
\end{eqnarray}
The upon mass matrix can be diagonalized by the $2\times 2 $ unitary transformation with mass eigenvalues
\begin{eqnarray}
m_{1,2} = {v_s \over 2 \sqrt{2}} \left [  \kappa_1 + \kappa_2 \pm \sqrt{(\kappa_1 - \kappa_2)^2  +8 \left({m\over v_s } \right)^2 } \right ] \; .
\end{eqnarray}
The corresponding eigenvectors are
\begin{eqnarray}
\hat \chi_1^{} &=& \cos \theta (\chi_L^{} +\chi_L^C) -  \sin \theta (\chi_R^{} + \chi_R^C) \;  ;\\
\hat \chi_2^{} &=&\cos \theta (\chi_R^{} + \chi_R^C) + \sin\theta (\chi_L^{} + \chi_L^C) \; ,
\end{eqnarray}
where $\theta$ is the mixing angle with $\tan^{-1} 2\theta=(\kappa_4^{}-\kappa_5^{})/2\kappa_6^{} $.
In this paper we take $\hat \chi_1$ as the dark matter candidate. 

The dark matter couples to $\mathbf{Z}'$ as well as CP-even scalars. 
Considering that the scalar mediated annihilation process is suppressed by mixing angles, we will mainly discuss the contribution of $\mathbf{Z}'$. 
The annihilation cross sections mediated by scalars are given in the appendix for completeness. 
The thermal average of the reduced annihilation cross sections take the form:
\begin{eqnarray}
\langle \sigma_{\overline{\hat \chi_1^{}} \hat \chi_1^{} \to e^+e^-  }  v \rangle &=& { g_\mathbf{X}^4  m^2   \over 12 \pi  [ (4m^2 -M_{\mathbf{Z'}}^2)^2 +M_{\mathbf{Z'}}^2 \Gamma^2_{\mathbf{Z'}} ]}  \langle v^2 \rangle \\
\langle \sigma_{\overline{\hat \chi_1^{}} \hat \chi_1^{} \to \mathbf{Z'}\mathbf{Z'}  }  v \rangle &=&  {g_\mathbf{X}^4 \over 16 \pi m^2 } \left( 1- {M_{\mathbf{Z}'}^2 \over m^2 } \right)^{3/2}  \left( 1- {M_{\mathbf{Z}'}^2 \over 2m^2 } \right)^{-2}  +  { \langle v^2\rangle  g_\mathbf{X}^4  \over 256 \pi M_{\mathbf{Z}'}^2} \left( 1- {M_{\mathbf{Z}'}^2 \over m^2 } \right)^{1/2}   \nonumber \\
&&   \left( 1- {M_{\mathbf{Z}'}^2 \over 2m^2 } \right)^{-4} \left( 128 { m^2 \over M_{\mathbf{Z'}}^2 } + 23 {M_{\mathbf{Z'}}^8  \over m^8 } -118 {M_{\mathbf{Z'}}^6  \over m^6 } + 172  {M_{\mathbf{Z'}}^4  \over m^4 }+  \right. \nonumber \\ && \left. 32  {M_{\mathbf{Z'}}^2  \over m^2 } -192  \right) \; .
\end{eqnarray}
where we have neglected the electron mass.

\begin{figure*}[t]
\includegraphics[width=0.49\textwidth]{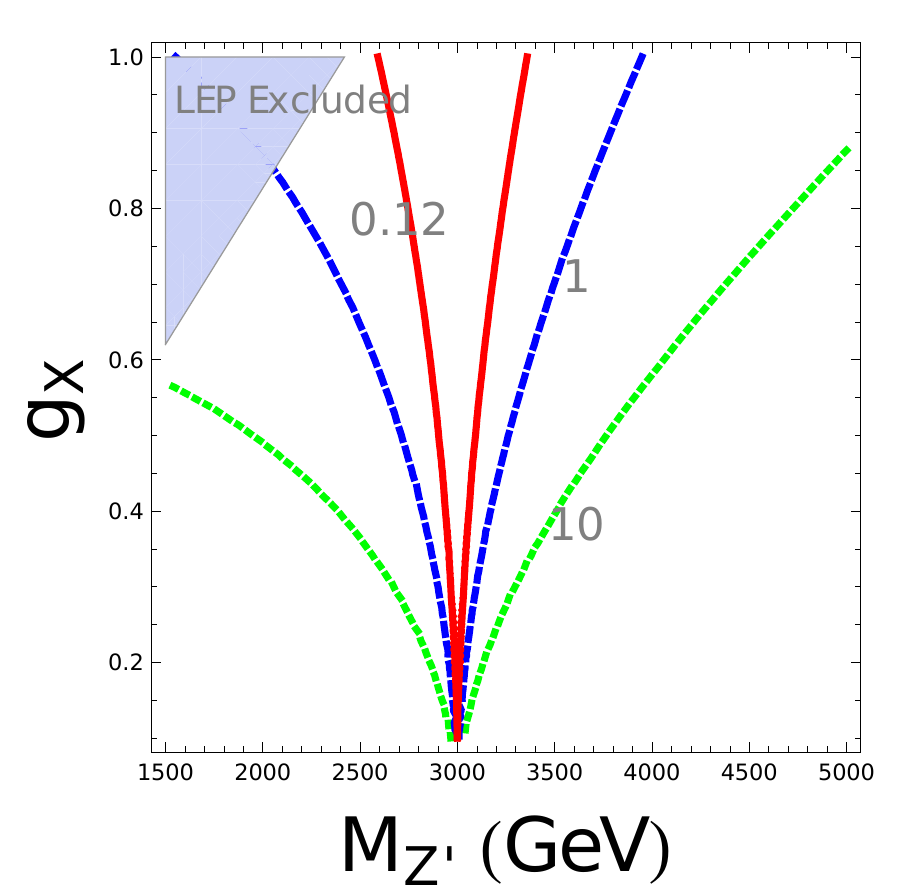}
\includegraphics[width=0.49\textwidth]{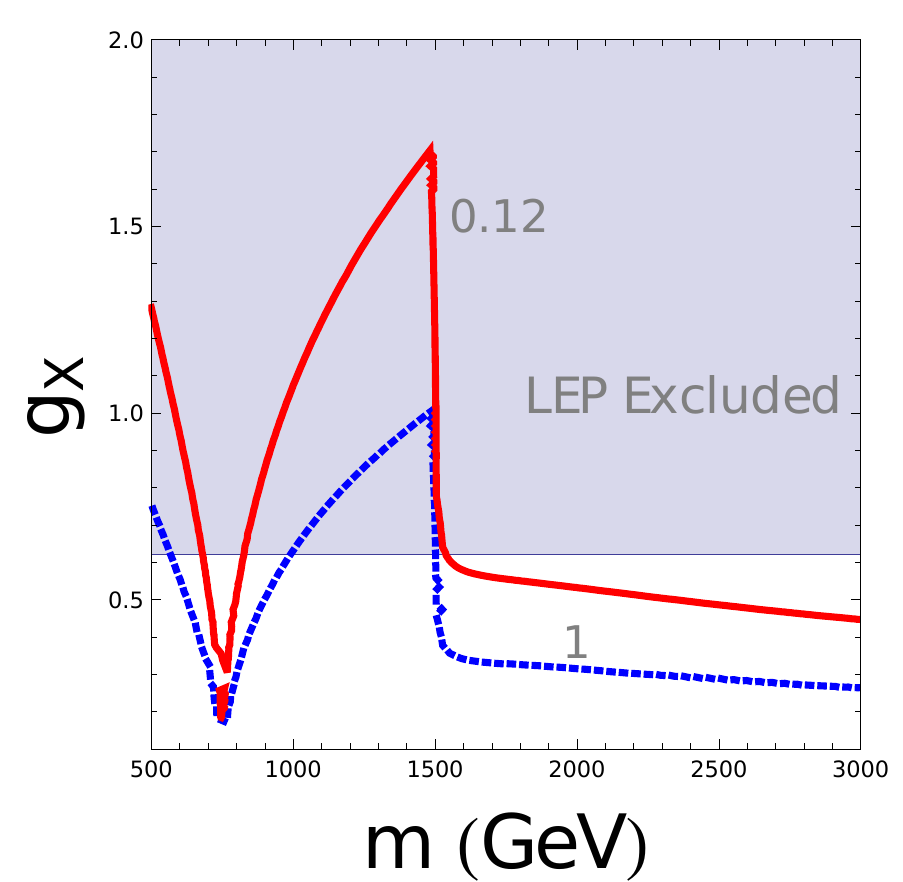}
\caption{Left-panel: contours of the Majorana dark matter relic density in the $M_{\mathbf{Z}'}-g_\mathbf{X}$ plane by setting $m=1.5~\text{TeV}$, the region in the top-left is excluded by the LEP; Right-panel: contours of the Majorana dark matter relic density in the $m-g_\mathbf{X}$ plane by setting $M_{\mathbf{Z}'}=1.5~\text{TeV}$, the region above the horizontal line  is excluded by the LEP.
\label{fig:relicB}}
\end{figure*}

We show in the Fig.~\ref{fig:relicB}, contours of the dark matter relic density  in the  $M_{\mathbf{Z}'}-g_\mathbf{X}$ plane by setting $m=1.5~\text{TeV}$(left-panel) and in the $m-g_\mathbf{X}$ plane by setting $M_{\mathbf{Z}'}=1.5~\text{TeV}$. 
The shaded region is excluded by the LEP.
The red solid line corresponds to $\Omega h^2 =0.12$ in both plots. 
One can conclude that both Dirac and Majorana dark matter scenarios can fit the observed relic density. 
For the direct detection, one has $\sigma^{\rm SI}\sim10^{-47} ~{\rm cm^2}$ by setting $s_\alpha \sim 0.01$, $m\sim 1.5~{\rm TeV}$ and $\xi\sim0.5$ (the Yukawa coupling in Eq.(\ref{a11}) ), where $\sigma^{\rm SI}$ is the spin independent DM-nucleon scattering cross section.
This value lies far below the current exclusion limits of PandaX-II~\cite{Cui:2017nnn} and XENON1T~\cite{Aprile:2017iyp}.

\section{The dampe cosmic ray excess}

\subsection{Dark matter distribution}

To reproduce the sharp feature of the DAMPE result, the annihilation 
of dark matter should dominantly occur in the local region of the 
solar system. A proper scenario is a relatively massive subhalo close 
to the Earth~\cite{Yuan:2017}. It has been expected that large amount 
of subhaloes exist in the Milky Way from high resolution numerical
simulations~\cite{Diemand:2005vz,Springel:2008cc}. The masses of 
subhalos can be as high as $10^{10}$ M$_{\odot}$, and as light as
the Earth mass~\cite{Diemand:2005vz}. In order that the spectrum of 
CREs will not get smoothed out by the radiative cooling during the 
propagation, we further require that the distance to the subhalo center 
is $\lesssim 0.3$ kpc~\cite{Yuan:2017}. 

The density profile of the subhalo is assumed to be a Navarro-Frenk-White 
profile~\cite{Navarro:1996gj}. Note that such a subhalo would be strongly 
affected by the tidal force of the Milky Way halo. The tidal radius is
approximately calculated as the radius at which the density of the subhalo 
is 0.02 times of the local average unbound density~\cite{Springel:2008cc}. 
The dark matter beyond the tidal radius is removed by the tidal force.
See Refs.~\cite{Springel:2008cc,Yuan:2017} for more details.

\subsection{CRE propagation}

The propagation of CREs in the Milky Way can be described by the
diffusion equation with energy losses
\begin{equation}
\frac{\partial \psi}{\partial t}=\nabla\cdot(D\nabla\psi)+
\frac{\partial}{\partial E}(b\psi)+q_e,\label{eq:prop-e}
\end{equation}
where $\psi$ is the differential number density of particles, $D\equiv D(E)$ 
is the diffusion coefficient which is assumed to be a function of energy, 
$b\equiv-{\rm d}E/{\rm d}t$ is the energy loss rate, and $q_e \equiv 
q_e(\boldsymbol{r},E)$ is the source injection function.

The cooling rate of CREs can be approximated as~\cite{Atoyan:1995ux}
\begin{equation}
b(E)=b_0+b_1E_{\rm GeV}+b_2E_{\rm GeV}^2,
\label{be}
\end{equation}
where $E_{\rm GeV}\equiv E/{\rm GeV}$. The three terms in the right hand
side represent the ionization, bremsstrahlung, and synchrotron and inverse
Compton energy losses, respectively. For the local environment with (neutral) 
gas density of $\sim1$ cm$^{-3}$ and magnetic field plus radiation field
energy density of $\sim1$ eV cm$^{-3}$, we have $b_0\approx3\times10^{-16}$ 
GeV s$^{-1}$, $b_1\approx10^{-15}$ GeV s$^{-1}$, and $b_2\approx1.0\times
10^{-16}$ GeV s$^{-1}$.

Assuming a spherically symmetric geometry with infinite boundary, which
is suitable for the energy range relevant in this work (TeV), the Green's 
function of Eq.~(\ref{eq:prop-e}) with respect to $r$ and $t$
reads~\cite{Atoyan:1995ux} 
\begin{equation}
{\mathcal G}(r,E,t)=\frac{N_{\rm inj}(E_i)b(E_i)}{\pi^{3/2}b(E)\lambda^3}
\exp\left(-\frac{r^2}{\lambda^2}\right). \label{eq:propagator}
\end{equation}
Here $N_{\rm inj}$ is the injection energy spectrum of CREs, $E_i$ is 
the initial energy of an electron which is cooled down to $E$ within 
time $t$, and 
\begin{equation}
\lambda(E)=2\left(\int_E^{\infty}\frac{D(E')}{b(E')}dE'\right)^{1/2},
\end{equation}
is the effective propagation length of CREs within the cooling time.
The convolution of Eq. (\ref{eq:propagator}) with the source spatial 
distribution and injection history gives the propagated electron fluxes 
at the Earth's location. In this work the diffusion coefficient is adopted 
to be $D(E)=D_0(E/{\rm GeV})^{\delta}$ with $D_0=3.3\times10^{28}$ 
cm$^2$ s$^{-1}$ and $\delta=1/3$. 

\subsection{Background}

In order to compare with the data, we also need to specify the background 
component. The background is assumed to be a double-broken power-law 
form as~\cite{Fan:2017}
\begin{equation}
\Phi_{\rm bkg}=\Phi_0\,E^{-\gamma}\left[1+\left(\frac{E_{\rm br,1}}{E}
\right)^{10}\right]^{\Delta\gamma_1/10}\,
\left[1+\left(\frac{E}{E_{\rm br,2}}\right)^{10}\right]
^{\Delta\gamma_2/10},
\end{equation}
with two breaks at $E_{\rm br,1}\sim50$ GeV and $E_{\rm br,2}\sim900$ 
GeV as revealed by the Fermi-LAT \cite{Abdollahi:2017nat} and DAMPE 
\cite{dampe-data} data. This complicated form may be due to the discrete
distribution of astrophysical CRE sources~\cite{DiMauro:2014iia,Fang:2016wid}. 
Such a background term can describe properly the overall behavior of the 
CRE fluxes from 10 GeV to 5 TeV~\cite{Fan:2017}. The parameters are
determined through a global fit to the DAMPE data including the dark
matter component (see below).

\subsection{Results}

\begin{figure}[t]
\includegraphics[width=0.7\textwidth]{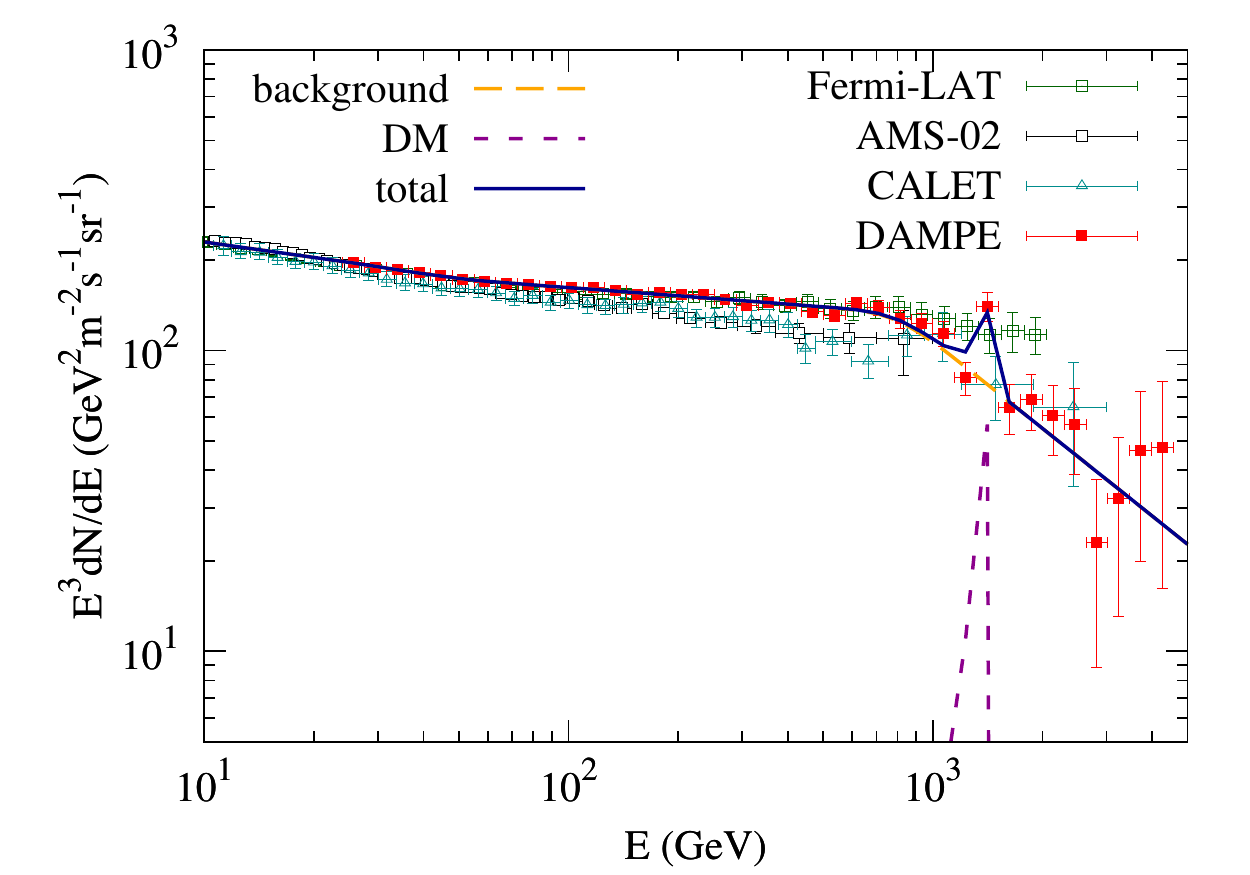}
\caption{Model prediction of the CRE fluxes from background (long-dashed),
DM annihilation (short-dashed), and the sum (solid), compared with the
DAMPE data \cite{dampe-data}. Also shown are the observational data from
Fermi-LAT \cite{Abdollahi:2017nat}, AMS-02 \cite{Aguilar:2014fea}, 
and CALET \cite{Adriani:2017efm}.
\label{fig:dampe_elec_flux}
}
\end{figure}

\begin{figure*}[t]
\includegraphics[width=0.49\textwidth]{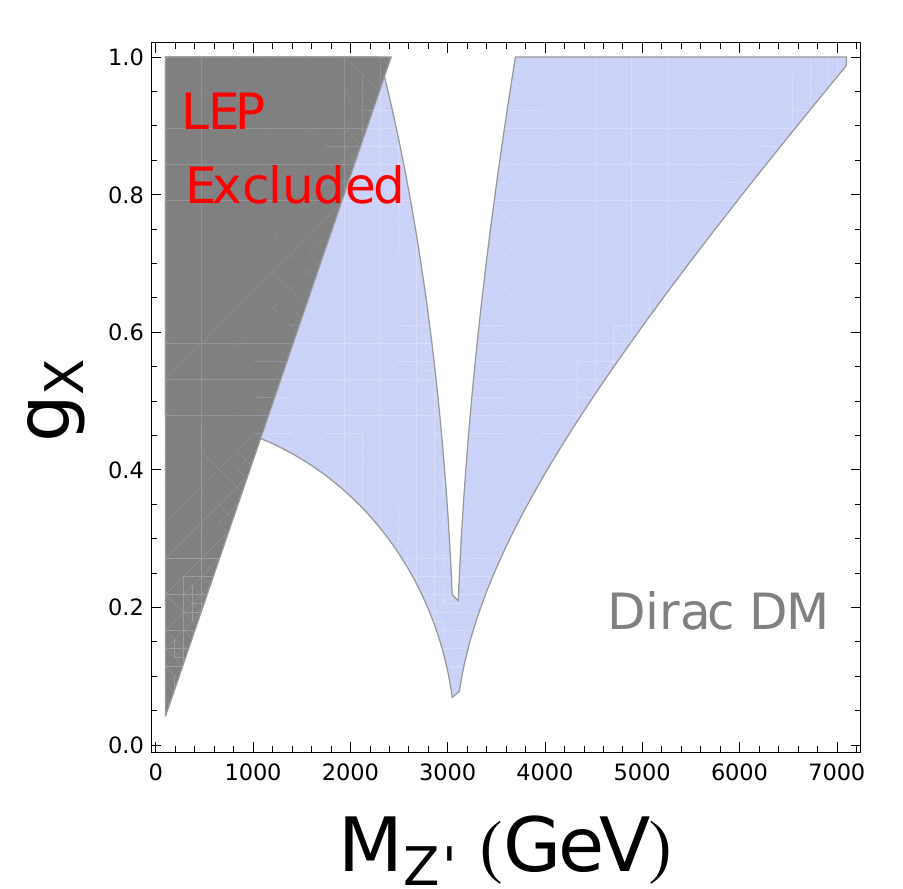}
\includegraphics[width=0.49\textwidth]{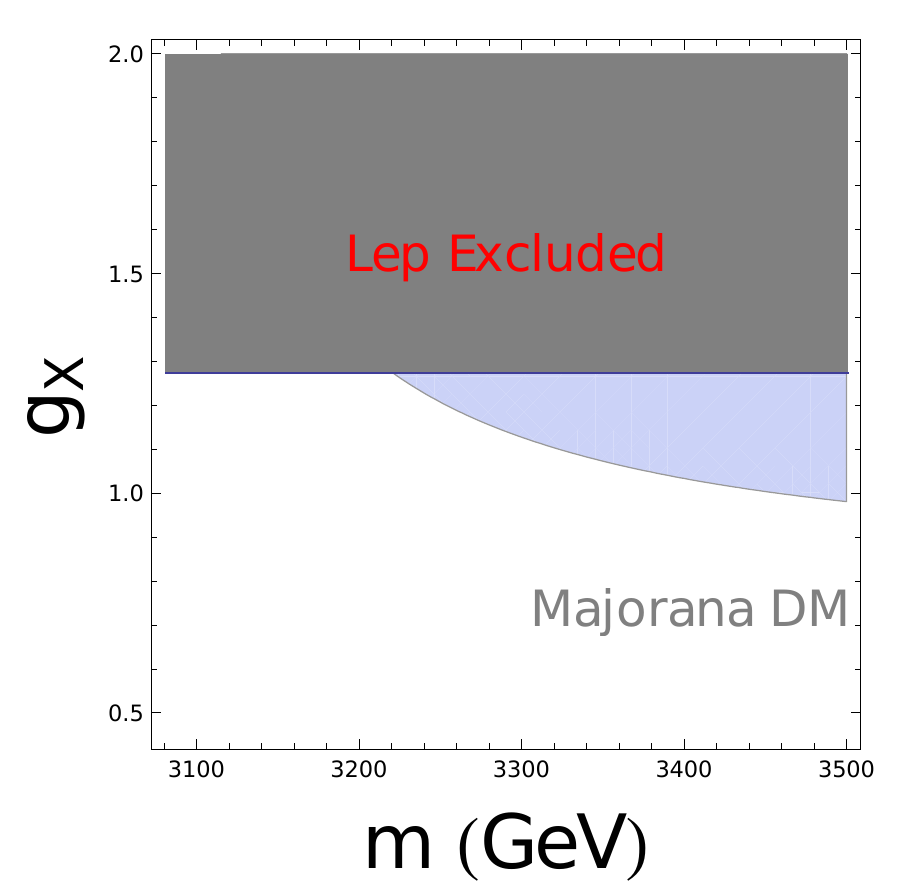}
\caption{Left-panel: parameter space in the $m_{\mathbf{Z}'}-g_\mathbf{X}$ region that can explain the DAMPE cosmic ray excess for the Dirac dark matter case, the gray region is excluded by the LEP; Right-panel: parameter space in the $m-g_\mathbf{X}$ plane that has  $\langle \sigma v\rangle \in (0.5\times 10^{-26} {\rm cm^3/s},~0.5\times 10^{-24} {\rm cm^3/s})$, the gray region  is excluded by the LEP.
\label{fig:dampep}}
\end{figure*}

The fit to the DAMPE data with the background and dark matter contribution
gives $\Phi_0=247.2$ GeV$^{-1}$~m$^{-2}$~s$^{-1}$~sr$^{-1}$, $\gamma=3.092$,
$\Delta\gamma_1=0.096$, $\Delta\gamma_2=-0.876$, and $E_{\rm br,2}=798.9$ 
GeV. The parameter $E_{\rm br,1}$ is not well determined by the DAMPE data
alone, and we fix it to be $50$ GeV~\cite{Abdollahi:2017nat}. It has little 
impact on the dark matter component around TeV scale discussed here. 
Fig.~\ref{fig:dampe_elec_flux} shows the best-fit result of the CRE fluxes 
compared with the data. Here we assume the mass of the subhalo is about
$10^6$ M$_\odot$, and the distance to the subhalo center is 0.1 kpc.
The mass of the dark matter particle is determined to be 1.54 TeV, and the 
annihilation cross section $\langle \sigma_{\bar{\chi}\chi\to \bar{e}e} 
v\rangle=3\times10^{-25}$ cm$^3$ s$^{-1}$, for the Dirac fermion scenario.
This set of model parameters gives a very good fit to the DAMPE data.
The cross section is degenerate with the mass and/or distance of the
subhalo. Considering the constraints from the relic density, the cosmic 
microwave background, and/or $\gamma$-rays, the cross section is limited
in a range of $10^{-26} \sim 10^{-24}$ cm$^3$ s$^{-1}$
\cite{Yuan:2017,Fan:2017}.
This requirement can be translated to constraints on the parameter space of our models, as are given in Fig.~\ref{fig:dampep}, where we show the parameter space in the $m_{\mathbf{Z}'}-g_\mathbf{X}$ plane that has $\langle \sigma v_{\chi \bar  \chi \to e\bar e} \rangle  \in  (10^{-26} {\rm cm^3/s},~10^{-24} {\rm cm^3/s})$ for the Dirac dark matter case (left-panel) and $\langle \sigma v_{\hat \chi_1 \hat \chi_1 \to \mathbf{Z}' \mathbf{Z}'}\rangle \in (0.5\times 10^{-26} {\rm cm^3/s},~0.5\times 10^{-24} {\rm cm^3/s})$ for the Majorana dark matter case (right-panel), the gray region  is excluded by the LEP.  
For the Dirac dark matter case, $\chi$ is fitted to be $1.54 ~\text{TeV}$ and the $Z^\prime$ should be roughly heavier than $1~{\rm TeV}$ taking into account the LEP constraint.  
For the Majorana dark matte case,  the $\mathbf{Z}'$ is fitted to be $3.08~\text{TeV}$ and the dark matter should be roughly heavier than $3.22~\text{TeV}$. 
Considering the shape of the observed cosmic ray excess, the Dirac dark matter case will be favored. 
It should be mentioned that  $\langle \sigma v_{\hat \chi_1 \hat \chi_1 \to \mathbf{Z}' \mathbf{Z}'}\rangle$ is further constrained by the CMB, FermiLAT and H.E.S.S, where the H.E.S.S. gives the most stringent constraint for $m>800~\text{GeV}$~\cite{Profumo:2017obk} and has $\langle \sigma v_{\hat \chi_1 \hat \chi_1 \to \mathbf{Z}' \mathbf{Z}'}\rangle<2\times 10^{-25} {\rm cm^3/s}$ for $m= 1~\text{TeV}$.

\section{Conclusion}

We have proposed  an electron flavored $\mathbf{Z}'$ portal dark matter model  to explain the observed excess. 
The model contains a vector-like charged singlet, a scalar singlet and a Dirac  or Majorana dark matter in addition to the SM  particles.
We showed that the constraint from the LEP is quite loose, which has $M_{\mathbf{Z}'}/g_\mathbf{X}\leq 2.42~{\rm TeV}$ and the electron mass may be generated from  a seesaw mechanism. 
To address the DAMPE cosmic ray excess, the annihilation of dark matter should dominantly occur in the local region of the solar system and the reduced cross section  is limited in a range of $10^{-26} \sim10^{-24} ~cm^3 s^{-1}$.
Our model can easily explain the observed relic abundance and the DAMPE cosmic ray excess while evading constraint of the direct detections for the Dirac dark matter case. 
It should be mentioned that this model may be tested by searching for extra scalar singlet in various  Di-Higgs channels, and the $\mathbf{Z}^\prime$ resonance in future $e^+e^-$ colliders, which, although interesting but beyond the reach of this paper,  will be given in another project.

\begin{acknowledgments}
This work was supported by the National Natural Science Foundation of China 
under Grant Nos. 11775025 and 11722328, the 100 Talents Program of 
Chinese Academy of Sciences and the Fundamental Research Funds for the Central Universities.
\end{acknowledgments}

\appendix
\section{Reduced annihilation cross section mediated by scalars}

Interactions between the Majorana dark matter and CP-even scalars can be written as 
\begin{eqnarray}
- {\cal L} \sim  { 1  \over 2 } \xi \overline{\hat \chi_1^{} } (c_\alpha s-s_\alpha h ) \hat \chi_1^{}  \label{a11}
\end{eqnarray}
where $\xi = (\kappa_1 c_\theta^2+\kappa_2 s_\theta^2)/\sqrt{2} $, $s$ and $h$ are mass eigenvectors of the CP-even scalars.

The thermal average of the reduced annihilation cross sections take the form: 
\begin{eqnarray}
&\langle \sigma v \rangle_{s_a s_b} =&  { \xi^2\over 1 + \delta } {\lambda^{1/2} (4, \boldsymbol{\zeta}_a^{}, \boldsymbol{\zeta}_b^{}) \over 1024 \pi v_\Phi^2 m_\mathbf{N}^2 } \left| {c_\alpha^{}  C_{s ab} \over 4-\boldsymbol{\zeta}_s^{}} -  { s_\alpha^{} C_{hab}\over 4 -\boldsymbol{\zeta}_h^{}} \right|^2 \langle v^2 \rangle   \\
%
%
%
%
&\langle \sigma v \rangle_{ZZ} = &
\frac{\xi^2 s_{2\alpha}^2 m_Z^4}{512\pi m^4 v^2 } \sqrt{1-\boldsymbol{\zeta}_Z^{}} \left ( 3-{4 \over \boldsymbol{\zeta}_Z^{}} + {4  \over \boldsymbol{\zeta}_Z^{2}}\right)
\left| { 1 \over 4-\boldsymbol{\zeta}_s^{}} -  { 1 \over 4 -\boldsymbol{\zeta}_h^{}} \right|^2
\langle v^2  \rangle,  \\
%
%
%
&\langle \sigma v \rangle_{WW} = & \frac{\xi^2 s_{2\alpha}^2 m_W^4}{256\pi m^2 v^2  } \sqrt{1-\boldsymbol{\zeta}_W^{}} \left ( 3-{4 \over \boldsymbol{\zeta}_W^{}} + { 4 \over \boldsymbol{\zeta}_W^{2}} \right)
\left| { 1 \over 4 -\boldsymbol{\zeta}_s^{}} -  { 1\over 4 -\boldsymbol{\zeta}_h^{}} \right|^2
\langle v^2  \rangle    \\
%
%
%
%
%
%
%
%
&\langle \sigma v\rangle_{f\bar f} = &   \frac{ \xi^2 n_C^f s_{2\alpha}^2 \boldsymbol{\zeta}_f^{}}{32 \pi   v^2 }(1- \boldsymbol{\zeta}_f^{})^{3/2}
 \left| { 1 \over 4 - \boldsymbol{\zeta}_s^{}} - { 1 \over 4 -\boldsymbol{\zeta}_h^{}} \right|^2 \langle v^2 \rangle 
 \end{eqnarray}
where $\lambda(x,~y,~z)\equiv x^2+y^2+z^2-2xy-2xz-2yz$, $\boldsymbol{\zeta}_X^{} =m_X^2 /m_\mathbf{N}^2 $,   $\delta_{ab}^{}=1$ (for $a=b$) and $0$ (for $a\neq b$);   $C_{s_i s_j s_k}$ are trilinear couplings; $s_{2\alpha} =\sin 2 \alpha$, $c_\alpha=\cos \alpha$, $n_C^f$ is the color factor of $f$. 
Explicit expressions of trilinear couplings can be found in the Table III of the Ref.~\cite{Chao:2016avy}.


\begin{thebibliography}{99}

\bibitem{Zwicky:1933gu} 
  F.~Zwicky,
  Helv.\ Phys.\ Acta {\bf 6}, 110 (1933)
  [Gen.\ Rel.\ Grav.\  {\bf 41}, 207 (2009)].
  
\bibitem{Ade:2015xua} 
  P.~A.~R.~Ade {\it et al.} [Planck Collaboration],
  Astron.\ Astrophys.\  {\bf 594}, A13 (2016)
  [arXiv:1502.01589 [astro-ph.CO]].
  
  
\bibitem{Aubourg:2014yra} 
  E.~Aubourg {\it et al.},
  Phys.\ Rev.\ D {\bf 92}, no. 12, 123516 (2015)
  [arXiv:1411.1074 [astro-ph.CO]].
  
  
\bibitem{Patrignani:2016xqp} 
  C.~Patrignani {\it et al.} [Particle Data Group],
  Chin.\ Phys.\ C {\bf 40}, no. 10, 100001 (2016).
  
  
\bibitem{ChangJin:550}
  J.~Chang, Chinese Journal of Space Science {\bf 34}, 550 (2014).

\bibitem{TheDAMPE:2017dtc} 
  J.~Chang {\it et al.} [DAMPE Collaboration],
  Astropart.\ Phys.\  {\bf 95}, 6 (2017)
  [arXiv:1706.08453 [astro-ph.IM]].

\bibitem{dampe-data}  DAMPE Collaboration, Nature in press, DOI:10.1038/nature24475.


\bibitem{Yuan:2017ysv} 
  Q.~Yuan {\it et al.},
  arXiv:1711.10989 [astro-ph.HE].
  
  
\bibitem{Fan:2017sor} 
  Y.~Z.~Fan, W.~C.~Huang, M.~Spinrath, Y.~L.~S.~Tsai and Q.~Yuan,
  arXiv:1711.10995 [hep-ph].
  


\bibitem{Fang:2017tvj} 
  K.~Fang, X.~J.~Bi and P.~F.~Yin,
  arXiv:1711.10996 [astro-ph.HE].
  
  
\bibitem{Duan:2017pkq} 
  G.~H.~Duan, L.~Feng, F.~Wang, L.~Wu, J.~M.~Yang and R.~Zheng,
  arXiv:1711.11012 [hep-ph].
  
\bibitem{Gu:2017gle} 
  P.~H.~Gu and X.~G.~He,
  arXiv:1711.11000 [hep-ph].
  
\bibitem{Zu:2017dzm} 
  L.~Zu, C.~Zhang, L.~Feng, Q.~Yuan and Y.~Z.~Fan,
  arXiv:1711.11052 [hep-ph].
  
\bibitem{Tang:2017lfb} 
  Y.~L.~Tang, L.~Wu, M.~Zhang and R.~Zheng,
  arXiv:1711.11058 [hep-ph].
  
\bibitem{Huang:2017egk} 
  X.~J.~Huang, Y.~L.~Wu, W.~H.~Zhang and Y.~F.~Zhou,
  arXiv:1712.00005 [astro-ph.HE].
  
\bibitem{Athron:2017drj} 
  P.~Athron, C.~Balazs, A.~Fowlie and Y.~Zhang,
  arXiv:1711.11376 [hep-ph].
  
\bibitem{Cao:2017ydw} 
  J.~Cao, L.~Feng, X.~Guo, L.~Shang, F.~Wang and P.~Wu,
  arXiv:1711.11452 [hep-ph].
  
\bibitem{Liu:2017rgs} 
  X.~Liu and Z.~Liu,
  arXiv:1711.11579 [hep-ph].

\bibitem{Chao:2017emq} 
  W.~Chao, H.~K.~Guo, H.~L.~Li and J.~Shu,
  arXiv:1712.00037 [hep-ph].
  
\bibitem{Duan:2017qwj} 
  G.~H.~Duan, X.~G.~He, L.~Wu and J.~M.~Yang,
  arXiv:1711.11563 [hep-ph].
  
\bibitem{Gu:2017bdw} 
  P.~H.~Gu,
  arXiv:1711.11333 [hep-ph].
  
  

















\bibitem{He:1991qd} 
  X.~G.~He, G.~C.~Joshi, H.~Lew and R.~R.~Volkas,
  Phys.\ Rev.\ D {\bf 44}, 2118 (1991).
  doi:10.1103/PhysRevD.44.2118


\bibitem{King:2005jy} 
  S.~F.~King, S.~Moretti and R.~Nevzorov,
  Phys.\ Rev.\ D {\bf 73}, 035009 (2006)
  doi:10.1103/PhysRevD.73.035009
  [hep-ph/0510419].



\bibitem{Mohapatra:1980qe} 
  R.~N.~Mohapatra and R.~E.~Marshak,
  Phys.\ Rev.\ Lett.\  {\bf 44}, 1316 (1980)
  Erratum: [Phys.\ Rev.\ Lett.\  {\bf 44}, 1643 (1980)].
  doi:10.1103/PhysRevLett.44.1316
  
  
\bibitem{Marshak:1979fm} 
  R.~E.~Marshak and R.~N.~Mohapatra,
  Phys.\ Lett.\  {\bf 91B}, 222 (1980).
  doi:10.1016/0370-2693(80)90436-0
  
  
\bibitem{Wetterich:1981bx} 
  C.~Wetterich,
  Nucl.\ Phys.\ B {\bf 187}, 343 (1981).
  doi:10.1016/0550-3213(81)90279-0
  
  
\bibitem{FileviezPerez:2010gw} 
  P.~Fileviez Perez and M.~B.~Wise,
  Phys.\ Rev.\ D {\bf 82}, 011901 (2010)
  Erratum: [Phys.\ Rev.\ D {\bf 82}, 079901 (2010)]
  doi:10.1103/PhysRevD.82.079901, 10.1103/PhysRevD.82.011901
  [arXiv:1002.1754 [hep-ph]].
  
\bibitem{Dulaney:2010dj} 
  T.~R.~Dulaney, P.~Fileviez Perez and M.~B.~Wise,
  Phys.\ Rev.\ D {\bf 83}, 023520 (2011)
  doi:10.1103/PhysRevD.83.023520
  [arXiv:1005.0617 [hep-ph]].
  
  
  
  
  
  
\bibitem{Chao:2010mp} 
  W.~Chao,
  Phys.\ Lett.\ B {\bf 695}, 157 (2011)
  doi:10.1016/j.physletb.2010.10.056
  [arXiv:1005.1024 [hep-ph]].


\bibitem{Chao:2015nsm} 
  W.~Chao,
  Phys.\ Rev.\ D {\bf 93}, no. 11, 115013 (2016)
  doi:10.1103/PhysRevD.93.115013
  [arXiv:1512.06297 [hep-ph]].
  
\bibitem{Chao:2016avy} 
  W.~Chao, H.~k.~Guo and Y.~Zhang,
  JHEP {\bf 1704}, 034 (2017)
  doi:10.1007/JHEP04(2017)034
  [arXiv:1604.01771 [hep-ph]].


\bibitem{Appelquist:2002mw} 
  T.~Appelquist, B.~A.~Dobrescu and A.~R.~Hopper,
  Phys.\ Rev.\ D {\bf 68}, 035012 (2003)
  doi:10.1103/PhysRevD.68.035012
  [hep-ph/0212073].
  
\bibitem{Ekstedt:2016wyi} 
  A.~Ekstedt, R.~Enberg, G.~Ingelman, J.~Löægren and T.~Mandal,
  JHEP {\bf 1611}, 071 (2016)
  doi:10.1007/JHEP11(2016)071
  [arXiv:1605.04855 [hep-ph]].

\bibitem{Chao:2017rwv} 
  W.~Chao,
  arXiv:1707.07858 [hep-ph].



\bibitem{Langacker:2008yv} 
  P.~Langacker,
  Rev.\ Mod.\ Phys.\  {\bf 81}, 1199 (2009)
  doi:10.1103/RevModPhys.81.1199
  [arXiv:0801.1345 [hep-ph]].





\bibitem{Alves:2013tqa} 
  A.~Alves, S.~Profumo and F.~S.~Queiroz,
  JHEP {\bf 1404}, 063 (2014)
  doi:10.1007/JHEP04(2014)063
  [arXiv:1312.5281 [hep-ph]].
  
\bibitem{Alves:2016cqf} 
  A.~Alves, G.~Arcadi, Y.~Mambrini, S.~Profumo and F.~S.~Queiroz,
  JHEP {\bf 1704}, 164 (2017)
  doi:10.1007/JHEP04(2017)164
  [arXiv:1612.07282 [hep-ph]].
  
\bibitem{Arcadi:2017hfi} 
  G.~Arcadi, M.~D.~Campos, M.~Lindner, A.~Masiero and F.~S.~Queiroz,
  arXiv:1708.00890 [hep-ph].

\bibitem{avector1}

S. L. Adler, Phys. Rev. {\bf 177}, 2426(1969).

\bibitem{avector2}

J. S. Bell and R. Jackiw, Nuovo Cimento A {\bf 60}, 47(1969).

\bibitem{avector3}

W. A. Barden, Phys. Rev. {\bf 184}, 1848(1969).

\bibitem{anog1}

R. Delbourgo and A. Salam, Phys. Lett. B {\bf 40}, 381(1972).

\bibitem{anog2}

T. Eguchi and P. G. O. Freund, Phys. Rev. Lett {\bf 37}, 1251(1976).

\bibitem{anog3}

L. Alvarez-Gaume and E. Witten, Nucl. Phys. B {\bf 234}, 269(1984).





\bibitem{Profumo:2014opa} 
  S.~Profumo, M.~J.~Ramsey-Musolf, C.~L.~Wainwright and P.~Winslow,
  Phys.\ Rev.\ D {\bf 91}, no. 3, 035018 (2015)
  doi:10.1103/PhysRevD.91.035018
  [arXiv:1407.5342 [hep-ph]].

\bibitem{Chao:2016vfq} 
  W.~Chao,
  arXiv:1601.06714 [hep-ph].

\bibitem{Chao:2016cea} 
  W.~Chao, M.~J.~Ramsey-Musolf and J.~H.~Yu,
  Phys.\ Rev.\ D {\bf 93}, no. 9, 095025 (2016)
  doi:10.1103/PhysRevD.93.095025
  [arXiv:1602.05192 [hep-ph]].
  
\bibitem{Eichten:1983hw} 
  E.~Eichten, K.~D.~Lane and M.~E.~Peskin,
  Phys.\ Rev.\ Lett.\  {\bf 50}, 811 (1983).

\bibitem{Schael:2013ita} 
  S.~Schael {\it et al.} [ALEPH and DELPHI and L3 and OPAL and LEP Electroweak Collaborations],
  Phys.\ Rept.\  {\bf 532}, 119 (2013)
  [arXiv:1302.3415 [hep-ex]].
  
\bibitem{Antusch:2006vwa} 
  S.~Antusch, C.~Biggio, E.~Fernandez-Martinez, M.~B.~Gavela and J.~Lopez-Pavon,
  JHEP {\bf 0610}, 084 (2006)
  doi:10.1088/1126-6708/2006/10/084
  [hep-ph/0607020].
  
  
\bibitem{Abada:2007ux} 
  A.~Abada, C.~Biggio, F.~Bonnet, M.~B.~Gavela and T.~Hambye,
  JHEP {\bf 0712}, 061 (2007)
  doi:10.1088/1126-6708/2007/12/061
  [arXiv:0707.4058 [hep-ph]].
  


\bibitem{Cui:2017nnn} 
  X.~Cui {\it et al.} [PandaX-II Collaboration],
  Phys.\ Rev.\ Lett.\  {\bf 119}, no. 18, 181302 (2017)
  [arXiv:1708.06917 [astro-ph.CO]].

\bibitem{Aprile:2017iyp} 
  E.~Aprile {\it et al.} [XENON Collaboration],
  Phys.\ Rev.\ Lett.\  {\bf 119}, no. 18, 181301 (2017)
  [arXiv:1705.06655 [astro-ph.CO]].

\bibitem{Yuan:2017} Q. Yuan et al., in preparation {\bf xxx}, xxx (2017).

\bibitem{Diemand:2005vz} 
  J.~Diemand, B.~Moore and J.~Stadel,
  Nature {\bf 433}, 389 (2005)
  [astro-ph/0501589].

\bibitem{Springel:2008cc} 
  V.~Springel {\it et al.},
  Mon.\ Not.\ Roy.\ Astron.\ Soc.\  {\bf 391}, 1685 (2008)
  [arXiv:0809.0898 [astro-ph]].

\bibitem{Navarro:1996gj} 
  J.~F.~Navarro, C.~S.~Frenk and S.~D.~M.~White,
  Astrophys.\ J.\  {\bf 490}, 493 (1997)
  [astro-ph/9611107].

\bibitem{Atoyan:1995ux} 
  A.~M.~Atoyan, F.~A.~Aharonian and H.~J.~Volk,
  Phys.\ Rev.\ D {\bf 52}, 3265 (1995).

\bibitem{Fan:2017} Y.-Z. Fan et al., in preparation {\bf xxx}, xxx (2017).

\bibitem{DiMauro:2014iia} 
  M.~Di Mauro, F.~Donato, N.~Fornengo, R.~Lineros and A.~Vittino,
  JCAP {\bf 1404}, 006 (2014)
  [arXiv:1402.0321 [astro-ph.HE]].

\bibitem{Fang:2016wid} 
  K.~Fang, B.~B.~Wang, X.~J.~Bi, S.~J.~Lin and P.~F.~Yin,
  Astrophys.\ J.\  {\bf 836}, no. 2, 172 (2017)
  [arXiv:1611.10292 [astro-ph.HE]].

\bibitem{Abdollahi:2017nat} 
  S.~Abdollahi {\it et al.} [Fermi-LAT Collaboration],
  Phys.\ Rev.\ D {\bf 95}, no. 8, 082007 (2017)
  [arXiv:1704.07195 [astro-ph.HE]].

\bibitem{Aguilar:2014fea} 
  M.~Aguilar {\it et al.} [AMS Collaboration],
  Phys.\ Rev.\ Lett.\  {\bf 113}, 221102 (2014).

\bibitem{Adriani:2017efm} 
  O.~Adriani {\it et al.},
  Phys.\ Rev.\ Lett.\  {\bf 119}, no. 18, 181101 (2017).

\bibitem{Profumo:2017obk} 
  S.~Profumo, F.~S.~Queiroz, J.~Silk and C.~Siqueira,
  arXiv:1711.03133 [hep-ph].
  
\end{thebibliography}
\end{document}